\documentstyle[emulateapj]{article}

\slugcomment{To appear in the Astrophysical Journal} 

\lefthead{Yun et al.}
\righthead{Radio Observations of High-z Dusty QSOs}

\def\et{{et al.}}
\newbox\grsign \setbox\grsign=\hbox{$>$} \newdimen\grdimen \grdimen=\ht\grsign
\newbox\simlessbox \newbox\simgreatbox
\setbox\simgreatbox=\hbox{\raise.5ex\hbox{$>$}\llap
     {\lower.5ex\hbox{$\sim$}}}\ht1=\grdimen\dp1=0pt
\setbox\simlessbox=\hbox{\raise.5ex\hbox{$<$}\llap
     {\lower.5ex\hbox{$\sim$}}}\ht2=\grdimen\dp2=0pt

\def\simless{\mathrel{\copy\simlessbox}}

\begin{document}
 
\title{Sensitive Radio Observations of High Redshift Dusty QSOs}
 
\author{M. S. Yun\altaffilmark{1}, C. L. Carilli\altaffilmark{1},
R. Kawabe\altaffilmark{2},  Y. Tutui\altaffilmark{3}, K. 
Kohno\altaffilmark{2}, \& K. Ohta\altaffilmark{4}}
 
\altaffiltext{1}{National Radio Astronomy Observatory, P.O. Box O, 
Socorro, NM, 87801.} 
\altaffiltext{2}{Nobeyama Radio Observatory, Nagano 384-1305, Japan.} 
\altaffiltext{3}{Institute of Astronomy, University of Tokyo, Mitaka, Tokyo 181-8588, Japan.} 
\altaffiltext{4}{Department of Astronomy, Kyoto University, Kyoto 606-8502, Japan.}

\setcounter{footnote}{4}

\begin{abstract}

We present sensitive radio continuum imaging at 1.4 GHz and 4.9 GHz
of seven high redshift QSOs selected for
having a 240 GHz continuum detection, which is thought to be thermal
dust emission.  We detect radio continuum emission from four of the
sources: BRI~0952$-$0115, BR~1202$-$0725, LBQS~1230+1627B, 
and BRI~1335$-$0417.  The
radio source in BR~1202$-$0725 is resolved into two components,
coincident with the double mm and CO sources.  We compare the results
at 1.4 GHz and 240 GHz to empirical and semi-analytic spectral models
based on star forming galaxies at low redshift. The radio-to-submm
spectral energy distribution for BR~1202$-$0725, LBQS~1230+1627B, and BRI~1335$-$0417 
are consistent with that expected for a massive starburst galaxy, with
implied massive star formation rates of order 10$^3$ M$_\odot$
year$^{-1}$ (without correcting for possible amplification by
gravitational lensing).  The radio-to-submm spectral energy distribution 
for BRI~0952$-$0115
suggests a low-luminosity radio jet source driven by the AGN.  

\end{abstract}
 

\section {Introduction}

Detecting submm thermal dust emission from objects at z $\ge$ 2
has revolutionized our understanding of galaxies at high redshift
(\cite{McMahon94,Hughes98,Ivison98a}, Smail, Ivison, \& Blain
1997, \cite{Barger98,Eales99}). 
A number of these submm sources have also been detected in 
CO emission with implied molecular gas masses $>$ 10$^{10}$ M$_\odot$ 
(\cite{Brown91,Barvainis94,Ohta96,Omont96a,Guilloteau97}, 
Frayer \et\ 1998, 1999), and the large reservoirs of warm gas and 
dust in these systems have led to the hypothesis that
these galaxies are starburst galaxies with massive star formation rates 
of order 10$^3$ M$_\odot$ year$^{-1}$ (Hughes \& Dunlop 1999, Lilly
\et\ 1999).

In their study of optically selected QSOs at z $>$ 4 from
the APM sample (McMahon 1991, Irwin, McMahon, \& Hazard 1991), Omont 
\et\ (1996b) detected 240 GHz continuum emission in 6 out of 16 
QSO's.  If the detected emission is due to dust in the 
quasar host systems, the implied dust masses are $\approx$ 10$^8$ 
M$_\odot$. Follow-up observations of these dust emitting QSOs have 
revealed CO emission in three cases so far, with implied 
molecular gas masses $\approx$ 10$^{11}$ M$_\odot$ 
(\cite{Ohta96,Omont96a}, Guilloteau et al. 1997,1999, Carilli, Menten,
\& Yun 1999). Omont \et\
also point out that about half the dust emitting QSOs have broad
absorption lines (BALs), again suggesting a gas rich environment.
Omont \et\ speculate that: ``...such large amounts of dust [and gas]
imply giant starbursts at z $>$ 4, at least comparable to those found
in the most hyperluminous IRAS galaxies...".  However, evidence
for active star formation in these sources remains circumstantial,
primarily based on the presence of large gas reservoirs, and it remains
possible that the dust is heated by the AGN, rather than by a starburst.

One possible method to investigate the nature of these high redshift
dust-rich QSOs is sensitive radio continuum observations.  A well
studied phenomenon in nearby star forming galaxies is the radio-to-far
IR correlation, i.e. the tight correlation found between radio
continuum emission and thermal dust emission (Condon 1992, Helou
\& Bicay 1993).  The standard explanation for the
radio-to-far IR correlation involves relativistic electrons
accelerated in supernova remnant shocks, and dust heated by the
interstellar radiation field.  Both quantities are then functions of
the massive star formation rate (Condon 1992, Cram \et\ 1998, 
Yun, Reddy, \& Condon 1999, Gruppioni, Mignal, \& Zamorani 1999), 
although the detailed physical processes giving rise to the tight 
correlation remain enigmatic.

An obvious difficulty in interpreting the radio and infrared
emission from an AGN host system in terms of 
star formation activity is disentangling the contribution
associated with the AGN activity.  A relatively minute amount
of dust ($10^{6-7} M_\odot$) heated by a luminous AGN can 
produce detectable levels of infrared and submm emission (e.g.
Yun \& Scoville 1998).  Similarly, all AGNs are likely sources of
radio emission at some level, and whether a particular QSO
is a radio source is really a question on sensitivity.
For example, a VLA 5 GHz radio continuum survey of 22
optically selected high redshift ($z>3$) QSOs by Schneider
et al. (1992) has produced only one detection above the
$5\sigma$ limit of 0.2 mJy\footnote{1 Jy $= 10^{-26}$ W
m$^{-2}$ Hz$^{-1}$} while another VLA survey of 
Palomar Bright Quasar Survey sources at $0.02<z<2.1$
by Kellermann et al. (1989) has detected 96 out of 114 (84\%)
at a similar limiting flux level.  The actual mechanism for
generating luminosity by an AGN is not understood well enough
nor particularly relevant to warrant further discussions here.
Given that radio luminosity of an AGN can span nearly ten
orders of magnitude, whether the observed radio luminosity of
a dusty QSO host system is even comparable to the value expected
from the inferred star forming activity is our main interest.
%In fact, it is shown that an elevation in radio luminosity 
%only 3-5 times the value inferred from the submm measurement
%can be easily recognized.

In this paper we present sensitive radio continuum observations of the
high redshift dust rich QSOs from the APM sample made with the 
Very Large Array (VLA) at 1.4 GHz and 4.9 GHz. 
For about half the sample, our observations are
sensitive enough to detect the radio continuum associated with a
starburst, as dictated by the dust continuum emission and the
radio-to-far IR correlation.  We find that three of the sources are
consistent with the radio-to-far IR correlation for starburst
galaxies while the radio emission in one of the sources is 
likely dominated by a low luminosity (FR~I)
radio jet driven by the AGN. We then examine whether other high redshift,
dust rich AGN hosts from the literature
also follow the radio-to-far IR correlation for
starburst galaxies, and we compare these results to the
low redshift ultraluminous IR galaxy and BAL QSO Mrk~231.
We adopt H$_\circ$ = 75 km s$^{-1}$ Mpc$^{-1}$ and q$_\circ$ = 0.5.

\section{Observations}

The sources listed in Table 1 are from a sample of optically
selected z $>$ 4 QSOs from the APM survey (\cite{Irwin91,McMahon91,Storrie96}).
They are selected for their very red color and point-like
appearance, with the idea that the red color is due to the
Ly$\alpha$ forest or the Lyman break shifting into the optical band.
An additional contribution to the red color may be dust obscuration
within the host galaxies.  Omont \et\ (1996b) observed 16
of these sources at 240 GHz to sensitivity limits of 1.5 mJy or
better using the bolometer array on IRAM 30-m telescope. 
They claim detections at $\ge$ 3$\sigma$ for six sources.  We
also include the z=2.70 QSO LBQS~1230+1627B, which was detected at 240
GHz by Omont \et\ among their low redshift weak emission line
sample.

Observations were made with the VLA in the BnA and C configurations 
at 1.4 GHz
and 4.9 GHz between 1998 and 1999.  Data from the VLA archive were
also examined in some cases to verify our results.  Integration times
were between 1 and 2 hours long, and a total bandwidth of 2 $\times$
50 MHz and two circular polarizations were used.  The absolute flux
density scale was set using 3C~286 (15 Jy 
at 1.4 GHz and 7.5 Jy at 4.9 GHz).  After normal
calibration using VLA calibrators, the data on each source were edited
and self-calibrated using standard methods for high dynamic range
imaging (Perley 1999).  Wide-field images encompassing the full
primary beam were synthesized and deconvolved in order to mitigate
image noise due to sidelobes from sources distributed throughout the
array field of view.  
The final rms noise levels on the images and measured source
flux density are listed in columns 5 and 6 in Table~1. In most cases
the final image noise values are within a factor 2 of the expected
theoretical limits.  The radio spectral index for each detected source
between 1.4 GHz and 4.9 GHz is listed in column~7 in Table~1.  We
define spectral index, $\alpha$, in terms of frequency, $\nu$, and the
observed flux density, S$_\nu$, as: S$_\nu$ $\propto$ $\nu^{\alpha}$.

%The column~8 in Table~1 
%lists the source flux densities at 350 GHz ($\lambda~850~\mu$m), extrapolated 
%from the 240 GHz flux density by Omont \et\ (1996b) assuming a spectral
%index of +3.25$\pm$0.25.  Column~8 lists the observed 1.4 GHz to 350
%GHz spectral index for each source or 4$\sigma$ lower limits.
%Column~10 lists the predicted value for a starburst galaxy using
%Eq.~3 of Carilli \& Yun (1999).  The
%uncertainty in the predicted values for $\alpha_{1.4}^{350}$ is at least
%0.1 due to the differences in the intrinsic properties of star forming
%galaxies (see Carilli \& Yun 1999).

\section{Results on Individual Sources}

\subsection{BRI~0952$-$0115}

This z = 4.43 QSO is the weakest of the claimed detections by Omont
\et\ (1996b) with S$_{240}$ = 2.8 $\pm$ 0.6 mJy.  Guilloteau \et\
(1999) have recently detected CO(5$-$4) emission from this QSO,
confirming its gas and dust rich nature.  The optical QSO is
gravitationally lensed with a 1$''$ separation (Omont \et\ 1996b), but
the radio and mm observations made thus far have not had sufficient
resolution to separate the two images.

This source is clearly detected at both 1.4 GHz and 4.9 GHz.  
%The radio source position (J2000) is: 09$^h$ 55$^m$ 00.12$^s$, 
%$-01^o$ 30$'$ 06.88$''$ ($\pm$ 0.5$''$).  
The source is not spatially resolved
by our observation at 4.5$''$$\times$2.7$''$ resolution, and the
measured flux density is 460$\pm$40 $\mu$Jy and 234$\pm$27 $\mu$Jy at
1.4 GHz and 4.9 GHz, respectively.  The radio spectral index between
1.4 and 4.9 GHz is $-0.54\pm 0.11$, which is slightly
flatter than the typical
value for a starburst system, $\alpha \approx -0.75 \pm0.1$ (Condon 1992).

If this were a starburst system at z = 4.43, then the expected 1.4 GHz
flux density extrapolated from the 240 GHz flux density would be 44
$\mu$Jy -- about 5 times smaller than the actual observed value.  
Similarly, the
observed value for $\alpha_{1.4}^{350}$ is +0.57$\pm$0.05 while the
predicted value for a starburst galaxy at that redshift is
+0.99$\pm$0.1.  Therefore this source has clear excess of radio emission
compared with a normal starburst galaxy. The radio luminosity of this
source is about 5 times that of the radio-loud elliptical galaxy M87,
and places this source in the class of low luminosity (Fanaroff-Riley
Class I) radio galaxies.  

One way to reconcile the starburst
hypothesis for driving the radio emission from this source is to
assume that gravitational lensing preferentially amplifies the radio
emission relative to the dust emission.  It is well known that
differential magnification by gravitational lensing can significantly
alter the apparent SED (e.g. Granato et al. 1996, Blain 1999a).
For a pure starburst system, however, the radio-to-submm flux 
ratio should be relatively unaffected by lensing
since radio synchrotron and thermal dust emission should have
similar spatial extents (\cite{Condon91}).  Gravitational
magnification is approximately inversely proportional to 
angular size, and a preferential amplification of radio emission
is expected only if a compact radio nucleus is present.
Several other sources discussed here may also be lensed, and
these considerations should generally apply to the others as well.

\subsection{BR~1202$-$0725}

The z = 4.70 QSO BR~1202$-$0725 is the first high redshift QSO where 
thermal dust emission was detected with S$_{240}$ = 12.6$\pm$2.3 
mJy and S$_{350}$ = 49$\pm$5 mJy, respectively
(McMahon \et\ 1994). This
source has also been detected in multiple transitions of CO (\cite{Ohta96,Omont96a,Yun99a}).
%, also revealing its dust and gas rich nature.  
This source has the curious property
that the optical QSO is a single source, but the mm continuum and the
CO line observations show a double source with a separation of about
4$''$ (\cite{Omont96a,Guilloteau99}). This double morphology may indicate 
a pair of interacting objects separated by only about 
20 kpc.  Alternatively, this may be a gravitationally lensed source with
one of the optical images obscured by intervening material (see below).

We have detected radio continuum emission from BR~1202$-$0725 with a
total flux density of 240$\pm$35 $\mu$Jy at 1.4 GHz. 
The low resolution image
(Figure~1a; $\theta = 10.6''\times6.6''$) shows a slight extension of
the source along the position angle of the mm continuum and CO
sources.  The high resolution image (Figure~1b; $\theta = 4.5''\times
2.8''$) shows a double source with the same separation and position
angle as the mm continuum and CO emitting sources. 
%The flux densities
%and J2000 positions of the two sources are: 70$\pm$35 $\mu$Jy at
%12$^h$ 05$^m$ 23.16$^s$, $-07^o$ 42$'$ 32.4$''$ and 133$\pm$35 $\mu$Jy
%at 12$^h$ 05$^m$ 23.02$^s$, $-07^o$ 42$'$ 29.8$''$ with position
%uncertainties of $\pm$ 1$''$. The flux density ratio between the two
%radio sources is 2$\pm$1.
This source is also detected at 4.9 GHz with a total flux
density of $141\pm15~\mu$Jy (Figure~1c).  The resulting 1.4 GHz to 4.9 GHz
spectral index is $-0.43\pm0.14$, which is flatter
than expected for a starburst galaxy.

The observed value of $\alpha_{1.4}^{350}$ for the integrated flux
density for BR~1202$-$0725 is +0.96$\pm$0.04, while the value predicted
for a starburst galaxy at z = 4.70 is +1.03$\pm$0.1.  Hence the 1.4
GHz-to-350 GHz spectral index is consistent with a massive starburst. 
This conclusion is somewhat inconsistent with the derived radio
spectral index above.  One possible explanation for this
discrepancy is free-free absorption of the radio continuum emission,
although the required emission measures are large 
($\approx$ few$\times 10^8$ pc cm$^{-6}$). Alternatively, 
a flat spectrum radio AGN may be present, contributing significantly
at 5 GHz, with steeper spectrum emission driven by star 
formation dominating at 1.4 GHz.  
This later alternative is admittedly {\sl ad hoc}, but
the situation is similar to what is seen in another AGN+starburst
system, Mrk~231 (see below). 
Future radio continuum imaging at sub-arcsecond resolution may help 
resolving this issue for BR~1202$-$0725. 

If the 1.4 GHz radio emission is driven by star formation, then the
required star formation rate is very large, $\approx$ 2500 M$_\odot$
year$^{-1}$ (using Eq.~1 of Carilli \& Yun 1999).  Again, 
given the differences among galaxies, a star
formation rate derived in this way should be considered an
order-of-magnitude estimate at best, and that the true star
formation rate could be a factor few to ten times lower if the source
is amplified by gravitational lensing.

BR~1202$-$0725 is found in a high density environment, 
and gravitational lensing may not be needed to explain the double
radio and submm morphology.    There is a Lyman-$\alpha$ 
emitting companion located $2.6''$ northwest of the 
quasar (Hu et al. 1996), and a deep ground based K-band image and an 
HST I-band image by Hu et al. show a string of continuum emission 
extending over 4$''$ in the same direction.  There is also a second 
galaxy located about $3.5''$ to the southwest of the quasar 
(see their Fig.~2).  One of the two radio continuum
sources is coincident with the optical quasar while the second 
radio and submm source is very close to but not coincident with the Lyman-$\alpha$ 
companion.  A possible faint K-band counterpart is visible in 
the image obtained by Hu et al., and the radio and submm source is 
probably also physically related to the Lyman-$\alpha$ source.  
Citing the close proximity ($\sim 20$ kpc), Hu et al. suggested 
ionization of the Lyman-$\alpha$ companion by the quasar.  However, 
external heating by the quasar can not explain the bright radio 
and submm emission in the companion.  If powered internally by 
star formation instead, the star formation rate inferred from the 
Lyman-$\alpha$ luminosity is only about 10$M_\odot$ yr$^{-1}$, which 
is about two orders of magnitudes smaller than inferred from the 
submm luminosity.  Similarly weak Lyman-$\alpha$ emission has been 
seen in other dusty high redshift submm sources such as [HR94] 10 
(ERO~J164502+4626.4, \cite{Hu94,Cimatti98a,Dey99}) and SMM~02399$-$0136 (\cite{Ivison98a}).  
The $\alpha_{1.4}^{350}$ index for the companion alone is 
+1.06$\pm$0.07, consistent with the SED of a z=4.7 starburst 
system.  In this scenario, the QSO is a flat spectrum
radio source located within a massive starburst host
while the companion is a second dust obscured starburst system.

\subsection{BRI~1335$-$0417}

The z = 4.40 QSO BRI~1335$-$0417 has been detected at 240 GHz by Omont
\et\ (1996b) with a flux density of 10.3$\pm$1.0 mJy and by
Guilloteau \et\ (1997) with a flux density of 5.6$\pm$1.1 mJy
at 220 GHz.  This
difference in observed flux densities at these two close frequencies
is consistent with a sharply rising submm spectral index 
(see Guilloteau \et\ 1997).  Given the large uncertainty in each
of these measurements, however, we adopt a mean value of 8.0$\pm$2.5 mJy at 240 
GHz with an uncertainty encompassing both measurements.  Then the
expected flux density at 350 GHz is 31$\pm$9 mJy.  This source has
also been detected in CO (5--4) emission by Guilloteau \et\ (1997), and
in the CO (2--1) transition by Carilli \et\ (1999),
with an inferred molecular gas mass of $\approx 10^{11} M_\odot$.
Neither the CO nor the thermal dust emission is resolved at 1$''$
resolution.

The source BRI~1335$-$0417 is detected in the radio continuum at 1.4 GHz
and 4.9 GHz with flux densities of 220$\pm$43 $\mu$Jy and
$76\pm11~\mu$Jy, respectively.  
%The J2000 position of the radio source peak is: 
%13$^h$ 38$^m$ 03.44$^s$, $-04^o$ 32$'$ 34.09$''$ ($\pm$ 2.0$''$).  
There is a possible confusing source about 15$''$ to the
east of BRI~1335$-$0417 (see Carilli \et\ 1999).  The 1.4 GHz to 4.9 GHz
spectral index is $-0.85\pm 0.19$, consistent with synchrotron
emission from a starburst system.  The observed value of
$\alpha_{1.4}^{350}$ for BRI~1335$-$0417 is +0.90$\pm$0.07 while the value
predicted for a starburst galaxy at z = 4.40 is +1.0$\pm$0.1.  Hence,
the radio synchrotron and mm dust emission from BRI~1335$-$0417 are
consistent with a massive starburst.  Like BR~1202$-$0725, the
required massive star formation rate is very large, $\approx$ 2000
M$_\odot$ year$^{-1}$.  Thus far there is no evidence for
gravitational lensing in BRI~1335$-$0417.

\subsection{LBQS~1230+1627B}

This z = 2.70 QSO is one of the low redshift QSOs in the Omont \et\ 
sample with a 240 GHz detection. Omont \et\ included this QSO in their 
search because they postulated that weak emission lines are  
indications of a possible dusty environment.  The reported 240 GHz 
flux is S$_{240}$ = 7.5$\pm$1.4 mJy.  The Plateau de Bure 
observations by Guilloteau et al. (1999) have confirmed this
detection, but they recovered only 3.3$\pm$0.5 mJy of flux
at 225 GHz.  The 850 GHz flux density of 
$104\pm21$ mJy has also been reported by Benford et al. (1999), and its 
dusty nature appears secure.  A previous search for radio continuum 
produced only a $3\sigma$ upper limit of 0.24 mJy at 8.4 GHz 
(\cite{Hooper95}). 

This source is detected at
both 1.4 GHz and 4.9 GHz with measured flux density of 
210$\pm$50 $\mu$Jy and 86$\pm$11 $\mu$Jy, respectively.
The derived radio spectral index $\alpha_{1.4}^{5} = -0.72\pm 0.22$ 
is typical of synchrotron emission from a star forming galaxy.
Adopting an average value of 5.0$\pm$1.5 mJy at 230 GHz, the derived $\alpha_{1.4}^{350}$ is +0.82$\pm$0.07, consistent within
the uncertainty with the predicted 
value for a starburst at z = 2.70 of  +0.71$\pm$0.10. 

\subsection{BR~1033$-$0327, BR~1117$-$1329, \& BR~1144$-$0723}

Along with BRI~0952$-$0115, these QSOs are the faintest sources detected by
Omont \et\ (1996b), with S$_{240}$ between 3.5 mJy and 5.8 mJy.  
The z = 4.51 QSO BR~1033$-$0327 was detected at 375 GHz
by Issak \et\ (1994) at 12$\pm$4 mJy while the extrapolated
350 GHz flux from the 240 GHz is 14$\pm$4 mJy.   The
extrapolated 350 GHz flux for the other two sources are 16 mJy 
and 23 mJy, respectively.  We do not detect these three 
sources at 1.4 GHz, with  4$\sigma$ upper limits
between 170 $\mu$Jy and 300 $\mu$Jy.  If the 240 GHz detection is
real, then the 4$\sigma$ lower limits to $\alpha_{1.4}^{350}$ are
$\approx$ +0.8 while the predicted values are +1.0$\pm$0.1.  Hence the
lower limits to $\alpha_{1.4}^{350}$ we obtained are consistent 
with a starburst interpretation, although these limits by no 
means preclude dust heated by an AGN in these systems.

\section{Discussion}

In the local universe, galaxies with a pronounced dust peak and large
FIR luminosity ($\ge 10^{12} L_\odot$) are luminous starburst systems
fueled by a massive concentration of gas, enshrouded in a thick dusty
cocoon (see Sanders \& Mirabel 1996 and references therein). The 240
GHz continuum emission detected by Omont \et\ in these high redshift
QSOs correspond to rest frame wavelengths of between 240 $\mu$m and
350 $\mu$m, near the peak of their FIR dust spectral energy
distribution (SED).  We first consider their large FIR luminosity in
relation to the ultraluminous infrared galaxies in the local universe.
Then we consider the clear presence of optical QSOs and the possible
AGN contribution to the radio and dust continuum emission.

\subsection{Starburst in Dusty QSOs}

The dust emission associated with these high redshift 
QSOs may be reprocessed radiation from luminous starburst
activity within their host galaxies is strongly suggested by
the comparison of their radio to infrared SEDs with that of
a starburst galaxy like M82 (see Figure~2).  While there is
a natural expectation of finding some radio continuum emission 
associated with AGN activity in these QSOs, detecting radio
emission exactly at the level expected from the dust emission
has to be entirely fortuitous.  The fact that this trend is
seen in three out of four QSOs detected at both radio and
mm/submm wavelengths is probably not coincidental.  Rather,
finding these luminous high redshift QSOs associated with
gas and dust-rich, massive starburst systems should have
been expected since most models of cosmic star formation 
history and galaxy formation/evolution trying to account for
the faint submm source counts require up to two orders of
magnitude increase in comoving density of such systems
at earlier epochs (\cite{Guiderdoni98,Blain99b,Tan99}).

The results of our radio continuum study of high redshift dust
emitting QSOs are summarized in Figure 3. This figure shows the values
of $\alpha_{1.4}^{350}$ versus redshift for the seven sources
discussed above. Also plotted are submm detected dust rich galaxies
with a wide range of redshifts and four models for the evolution of
$\alpha_{1.4}^{350}$ with redshift for starburst galaxies as given in
Carilli \& Yun (1999). Two of the models are semi-empirical, based on
starburst galaxy relationships derived by Condon (1992), while two of
the models are empirical, based on the observed spectra of the low
redshift star forming galaxies Arp~220 and M82.  
The radio-to-submm SEDs for the three sources,
BR~1202$-$0725, LBQS~1230+1627B, and BRI~1335$-$0417, are 
consistent, within the uncertainty and the range of the 
models, with starburst-driven radio
synchrotron and mm dust emission, despite the obvious presence of an
optical QSO. One source, BRI~0952$-$0115, shows clear
excess radio emission over the expected from a starburst.

Omont \et\ have claimed $\ge$ 3$\sigma$ detection of thermal dust emission at
240 GHz in 6 out of 16 QSOs from the APM sample at z $\ge$ 4.  Three of
these sources have also been detected in CO emission, and Omont \et\ point
out that half the sources are BAL QSOs -- a phenomenon that is
typically anti-correlated with radio-loud QSOs (see Weymann
1997). Overall, the host galaxies of these QSOs appear to be extremely
gas rich, leading Omont \et\ (1996a) to speculate that these AGN are
associated with starburst host galaxies.
However, detecting millimeter continuum emission alone does not serve
uniquely as an indicator of a gas- and dust-rich galaxy vigorously
forming stars since the presence of a flat spectrum AGN can account
for the mm/submm detection in some cases (e.g. B2~0902+34, see
\cite{Yun96,Downes96}).  Relatively small amount of hot dust 
($10^{6-7} M_\odot$) heated by an AGN can also produce elevated submm emission (\cite{Yun98,Haas98}).  The fact that six out of seven 240 GHz
detected QSOs have the same characteristic radio-to-submm
SED (including lower limits), consistent with that of a
starburst galaxy argues that these QSOs reside in gas and dust rich
host galaxies harboring a massive  starburst. The gas depletion 
time scale is short, $\approx$ 10$^{8}$ years, during which time 
a significant fraction of the stars in the host galaxy may be formed. 

There is a considerable scatter in the data and models plotted in
Figure~3, and this relation may only be used as gross redshift 
{\it indicator}.  Understanding the nature of contributing physical 
causes may allow a more robust use of this relation.  
An obvious concern is the contribution from a radio AGN (i.e. a jet), 
and the presence of an energetically important
AGN in general is discussed in greater detail below.
The intrinsic scatter in the radio-to-FIR correlation is only about
0.25 in dex (see Condon 1992), much smaller than the 
scatter seen in Figure~3.  The mean electron density, which is a
function of the compactness of the starburst region and the
intensity of the starburst activity, can produce a varying degree of 
free-free absorption of radio continuum (see Condon et al. 1991), and
this largely accounts for the systematic difference 
between the empherical models
based on ``M82" and ``Arp~220" in Figure~3.  As discussed by
Blain (1999b), dust temperature and emissivity can also introduce 
variations at the submm/FIR wavelengths that may contribute to
the observed scatter.  To produce a significant deviation from the
ranges of models considered in Figure~3, however, dust temperature has 
to be exceptionally low ($\simless 20$ K).  The measured
dust SEDs of infrared luminous galaxies in the local
universe are consistently characterized by dust temperature in excess
of 30 K (see Benford 1999, Lisenfeld et al. 1999), and the models
seem quite robust when applied to actively star forming galaxies.

It is well known that the fraction of radio loud QSOs decreases from
about 20$\%$ at z $\le$ 2 to 5$\%$ at z $\ge$ 3
(\cite{McMahon91,Schneider92,Schmidt95}).  In this case, ``radio-loud"
implies radio luminosity an order of magnitude or more larger than the
values presented herein, indicative of a radio jet associated
with the active nucleus.  The importance of the observations presented
here is that the flux density limits adequate to detect the
synchrotron radio continuum emission associated with star formation
%and not associated with a jet driven by the AGN, 
have been reached at least for some
sources. Hence we are potentially probing a new population of
radio-emitting AGN host galaxies at high redshift.
\bigskip

\subsection{Presence of a Luminous AGN}

There are now more than a dozen high redshift sources where  
thermal dust emission is detected.  Many of these sources
exhibiting a value of $\alpha_{1.4}^{350}$ characteristic of 
a starburst, and they also show clear evidence for an AGN. 
One such source is the z = 2.28 gravitationally 
lensed hyperluminous infrared galaxy IRAS 10214+4724.  Presence of a 
hidden QSO is evidenced by broad optical emission lines 
seen in scattered light (\cite{Lawrence93,Jannuzi94}).  
A second source is the z = 2.903 submm galaxy SMM~02399$-$0136, which 
is the first submm identified luminous galaxy showing an AGN spectrum 
(\cite{Ivison98a}).  The third is APM~08279+5255 at z = 3.87, which 
shows an optical BAL QSO spectrum (\cite{Irwin98,Lewis98,Hine99}). 
These sources have also been detected in CO emission and thus harbor 
large amounts of molecular gas (\cite{Brown91,Frayer98,Downes99}).  

The z = 2.558 BAL QSO H~1413+117 is another submm and CO source
(Barvainis \et\ 1992,1994), but its radio emission has a significant
contribution from its radio jet (\cite{Kayser90}).  Therefore it
deviates significantly from the radio-submm relation for starburst
galaxies as shown in Figure~3, similarly as BRI~0952$-$0115.  Other 
submm detected radio galaxies and radio-loud
QSOs such as 4C~41.17 at z = 3.80 (\cite{Dunlop94,Hughes97}), 8C~1435+635
at z = 4.25 (\cite{Ivison95,Ivison98b}), MG~0414+0534 at z = 2.64
(\cite{Barvainis98}), and MG~1019+0535 at z = 2.76 (\cite{Cimatti98b})
have comparable submm/IR luminosity, but their radio luminosity
is 2-3 orders of magnitude larger.  As a result, 
they deviate so much from the radio-submm relation as to drop
completely below the ranges plotted in Figure~3.

While we have argued for active star formation in at least some of
these systems, the question remains as to what degree the luminous
AGN present in these dusty QSOs account for the radio
and/or submm/IR luminosity.  The answer to this question
is difficult to obtain even for the ultraluminous infrared galaxies in the
local universe (see reviews by Sanders [1999], 
Joseph [1999], and Genzel [1999] and
references therein), and addressing this issue for the high redshift
objects has to be even harder.  Instead, an examination of a low redshift
analog of a galaxy with a concurrent AGN plus starburst in the
ultraluminous infrared galaxy Mrk~231 may provide a useful
insight.  Mrk~231 is the most luminous infrared source at z $\le$ 0.2
(Sanders \et\ 1988), with an infrared luminosity of $5\times10^{12}$
h$^{-2}$ L$_\odot$.  This is within a factor of a few to ten of the
high redshift dust emitting sources, and Mrk~231 has been called the
nearest BAL QSO (Forster \et\ 1995). About half of its 1.4 GHz radio
emission comes from a parsec-scale radio jet (Ulvestad \et\ 1999)
while the other half is associated with a gas disk seen in CO emission
and HI 21cm absorption on scales of 100 to 1000 pc
(\cite{Bryant96,Downes98,Carilli98}).  The thermal dust and
non-thermal radio continuum emission from this disk is consistent with
a massive starburst of 200 M$_\odot$ per year, and support for such a
star forming disk in Mrk~231 comes from the tentative detection of
supernova remnants (Taylor et al.  1999).  Citing geometric and
energetic reasons, Downes \& Solomon (1998) have argued that the AGN and the
starburst each account for about 1/2 of the total luminosity in
Mrk~231. The radio flux density of Mrk~231 between 5 GHz and 110 GHz 
is a factor 3 higher than the starburst prediction (see Figure~2)
due to the contribution from the radio jet.  This component is 
free-free absorbed at lower frequencies, where the flux densities 
are more in-line with those expected for a starburst.

Mrk~231 is an illustrative example of a dusty QSO where an AGN
may provide significant infrared and radio luminosity.  It is 
the closest local analog of the dusty high redshift QSOs 
whose radio emission has been studied here, but it is not
certain if Mrk~231 is representative of the dusty QSOs in
the early epochs.  Nevertheless,
the properties of Mrk~231 are consistent with the sequence proposed by
Sanders \et\ (1988) in which merging galaxies evolve through a
starburst phase, eventually becoming optical QSOs (see also Sanders \&
Mirabel 1996). The time scale for such evolution is of order a
few$\times$10$^8$ to 10$^9$ years. Sanders \et\ propose that Mrk~231
represents a late stage in this process, namely the `dust enshrouded
QSO' stage. Based on their radio and mm properties, many dust emitting
high redshift QSOs also appear to fall in this category although the
required star formation rates are a factor of 5 to 10 larger than IR
luminous galaxies at low redshift (in the absence of amplification by
gravitational lensing).  Clearly, deeper radio and mm observations at
sub-arcsecond spatial resolution are required to test this
hypothesis. Such observations will become possible with the upgraded
VLA at cm wavelengths and with the future Atacama Large Millimeter Array in Chile.

Despite the large error bars associated with the data points 
and a significant spread among the models, 
the dusty AGNs (filled and empty circles) as a group may suggest a 
systematic deviation by about 0.2-0.3 from the radio-submm 
relation obeyed by the pure starburst systems (empty squares) 
in Figure~3.  It is tempting to interpret this trend as a 
residual effect of energy input by an energetic AGN in
some cases, similarly as in Mrk~231 and to a lesser 
degree as in H~1413+117 and BR~0952$-$0115.  
Better and additional data on the high redshift dust-rich
sources identified by SCUBA as well as more refined models 
with improved understanding of the intrinsic scatter
are needed to examine the validity of this possible trend in detail.

\section{Summary}

The 1.4 and 4.9 GHz radio continuum emission is detected in four out of seven
high redshift QSOs that had previously be detected in thermal dust
emission at 240 GHz by Omont \et\ (1996b).  In three of the cases, the
cm and mm emission are consistent with starburst activity.  One of
the QSOs clearly hosts a radio AGN.  If the detected radio continuum 
is interpreted solely in terms of AGN activity instead, the
observed luminosity being within a factor of 2 of the value inferred
for starburst activity would be fortuitous, although such a
possibility cannot be ruled out.  

\vskip 0.2truein 

The authors thank F. Owen, A. Blain, I. Robson, R. Ivison, and 
K. Menten for useful
discussions. This research made use of the NASA/IPAC Extragalactic
Data Base (NED) which is operated by the Jet propulsion Lab, Caltech,
under contract with NASA. The National Radio Astronomy Observatory
(NRAO) is a facility of the National Science Foundation, operated
under cooperative agreement by Associated Universities, Inc.

\vfill\eject

\vfill\eject

%\centerline{Figure Caption}

\figcaption{
(a) A low resolution 1.4 GHz image of the z = 4.70
QSO BR~1202$-$0725 made with the VLA. The angular resolution is
10.6$''$$\times$6.6$''$ (FWHM, with major axis 
position angle = 90$^o$).  The crosses show the positions of the mm
continuum and CO sources by Omont \et\ (1996b). 
The contour levels are: $-$70, 70, 105, 140, 175, 210, and 245 $\mu$Jy
beam$^{-1}$. 
(b) The same as 1a, but now at higher resolution:
4.5$''\times 2.8''$ (FWHM) with major axis position angle = 82$^\circ$. 
The contour levels are $-$64, $-$32, 32, 64, 96, and 128 $\mu$Jy beam$^{-1}$.
The emission appears to be resolved into two separate sources roughly 
coincident with the positions of the two submm sources reported by Omont 
et al.  The flux densities
and J2000 positions of the two sources are: 70$\pm$35 $\mu$Jy at
12$^h$ 05$^m$ 23.16$^s$, $-07^o$ 42$'$ 32.4$''$ and 133$\pm$35 $\mu$Jy
at 12$^h$ 05$^m$ 23.02$^s$, $-07^o$ 42$'$ 29.8$''$ with position
uncertainties of $\pm$ 1$''$. 
%The flux density ratio between the two radio sources is 2$\pm$1.
(c) A 4.9 GHz image of the z = 4.70 QSO BR~1202$-$0725 
made with the VLA.  The angular resolution is
16.2$''$$\times$7.2$''$ (FWHM, with major axis 
position angle = $-$69$^o$). 
The contour levels are: $-$45, $-$30, 30, 45, 60, 75, 90, 105, 120,
and 135 $\mu$Jy beam$^{-1}$. \label{fig1}}

\figcaption{The spectral energy distribution of 1335$-$0415 (squares)
and BR~1202$-$0725 (circles) are compared with that of the 
prototypical starburst galaxy M82.  The vertical
scale is set for the flux density of M82, and the SEDs for all other
sources are normalized to M82 at the peak of their dust SED.  
The intrinsic 1.4 GHz luminosity (i.e. in the source rest frame) of 
BRI~1335$-$0415 and BR~1202$-$0725 are a factor 1500 or so higher 
than M82.  The SED for Mrk~231, an ultraluminous infrared galaxy and
the nearest BAL QSO, is shown (stars with an error bar) as
an illustrative example of a dusty QSO hosting a radio AGN.  It
has about 100 times larger 1.4 GHz luminosity than M82, and
its flat spectrum radio nucleus and a radio
jet accounts for a systematic offset between $10^{10-11}$ Hz.
%The data points between 
%5 GHz and 110 GHz are a factor 3 higher than the
%starburst prediction due to the contribution from the radio jet.
%seen on scales $\le$30 pc.  
%This component is free-free absorbed
%at lower frequencies, where the flux densities are more in-line with
%those expected for a starburst.   
\label{fig2}}

\figcaption{The data points show derived values of
$\alpha_{1.4}^{350}$ versus redshift for the sources in Table 1,
including 4$\sigma$ lower limits.  The lines are models derived for
starburst galaxies from Carilli \& Yun (1999).  Two power-law models
for $\alpha_{1.4}^{350}$ for star forming galaxies as a function of
redshift are shown with a solid curve ($\alpha_{\rm submm}=$ +3.0) and
a short dashed curve ($\alpha_{\rm submm}=$ +3.5).  The assumed radio
spectral index is $-$0.8.  The dotted and long-dashed curves show the
expected values of $\alpha_{1.4}^{350}$ versus redshift based on the
observed SED of the star forming galaxies M82 and Arp~220,
respectively.  The Omont \et\ sample QSOs and Mrk~231 are plotted as
filled circles while the lower limits for the three undetected Omont
\et\ QSOs are shown as arrows.  Dusty sources from the literature with
clear evidence for an AGN are also plotted as empty circles while the
sources without  evidence for an AGN are plotted as empty squares.
\label{fig3}}

\end{document}